\documentstyle[11pt,newpasp_crete2002,twoside,epsf]{article}
\begin{document}
\title{Deep Searches for Young Pulsars}
\author{Fernando Camilo}
\affil{Columbia University, 550 West 120th Street, New York, NY~10027, USA}

\begin{abstract}
By 2000 there were only 10 established Galactic pulsar--supernova
remnant associations.  Two years later there are 16 such associations
known.  I discuss the work leading to this substantial increase.  In
particular I summarize an ongoing search project that has resulted in
the detection of several young pulsars with a very low radio luminosity
of about 1\,mJy\,kpc$^2$ at 1.4\,GHz, and comment on future
prospects.
\end{abstract}

\vspace{-5mm}

\section{Introduction}
Young pulsars are objects of interest for a variety of reasons:
inferring the initial period, velocity, and magnetic field
distributions of neutron stars may inform the physics of core collapse;
measuring their ``beaming fraction'', luminosities, and spectra is
crucial for making a census of the Galactic population and determining
the birthrate of pulsars; also, they frequently exhibit period
glitches, and emit substantial amounts of X- and $\gamma$-rays ---
these can be observed to learn about the internal composition of
neutron stars, and their emission mechanisms.  In addition, many young
neutron stars are embedded in compact non-thermal radio and/or X-ray
pulsar wind nebulae (PWNe) where the ambient medium confines the
relativistic pulsar wind, or otherwise interact with their host
supernova remnants (SNRs); they are therefore unique probes of their
immediate environment and the local interstellar medium (ISM).

A natural location to search for young pulsars is the Galactic plane
and more specifically SNRs.  However, establishing bona fide
associations between Galactic pulsars and SNRs has been a painfully
slow-going business: two were known by 1970 (those of the Crab and
Vela), five by 1985, and only 10 by 2000 (see compilation by Kaspi \&
Helfand 2002), despite more than 200 SNRs and 1400 pulsars known.  It
is instructive to consider {\em how\/} those associations were
established: in four cases the pulsars were discovered in undirected
searches, even if the respective SNRs were already known
(B0531+21/Crab; B0833--45/Vela; B2334+61/G114.3+0.3; J0538+2817/S147);
in a further three cases the pulsars were detected in directed searches
of known point/X-ray sources (B1509--58/G320.4--1.2; B1951+32/CTB~80;
J1811--1925/G11.2--0.3); while in only three cases the pulsars were
discovered in unbiased searches of (sometimes large) SNRs
(B1757--24/G5.4--1.2; J1341--6220/G308.8--0.1; B1853+01/W44).

Clearly the wholesale search of SNRs has not been immensely productive
in this regard --- indeed, a survey of 88 (77 entire) SNRs in the 1990s
netted {\em zero\/} associated pulsars (Kaspi et al. 1996; Gorham et
al. 1996; Lorimer, Lyne, \& Camilo 1998).  One problem is that the
large total area to be searched limits the integration time used per
telescope pointing.  More recently the Parkes multibeam pulsar survey,
using a 13-beam receiver system at a frequency of 1.4\,GHz with a
bandwidth of 0.3\,GHz and individual 35\,min pointings, covered a very
large area ($|b|<5\deg$; $260\deg<l<50\deg$) with sensitivity broadly
comparable to that of the previous best SNR surveys.  While
extraordinarily successful, discovering more than 600 pulsars (e.g.,
Manchester et al.  2001), this survey to date has yielded only one new
pulsar--SNR association (see Fig.~1~left).  I now describe a far more
efficient method for detecting young pulsars.

\begin{figure}[h]
\plottwo{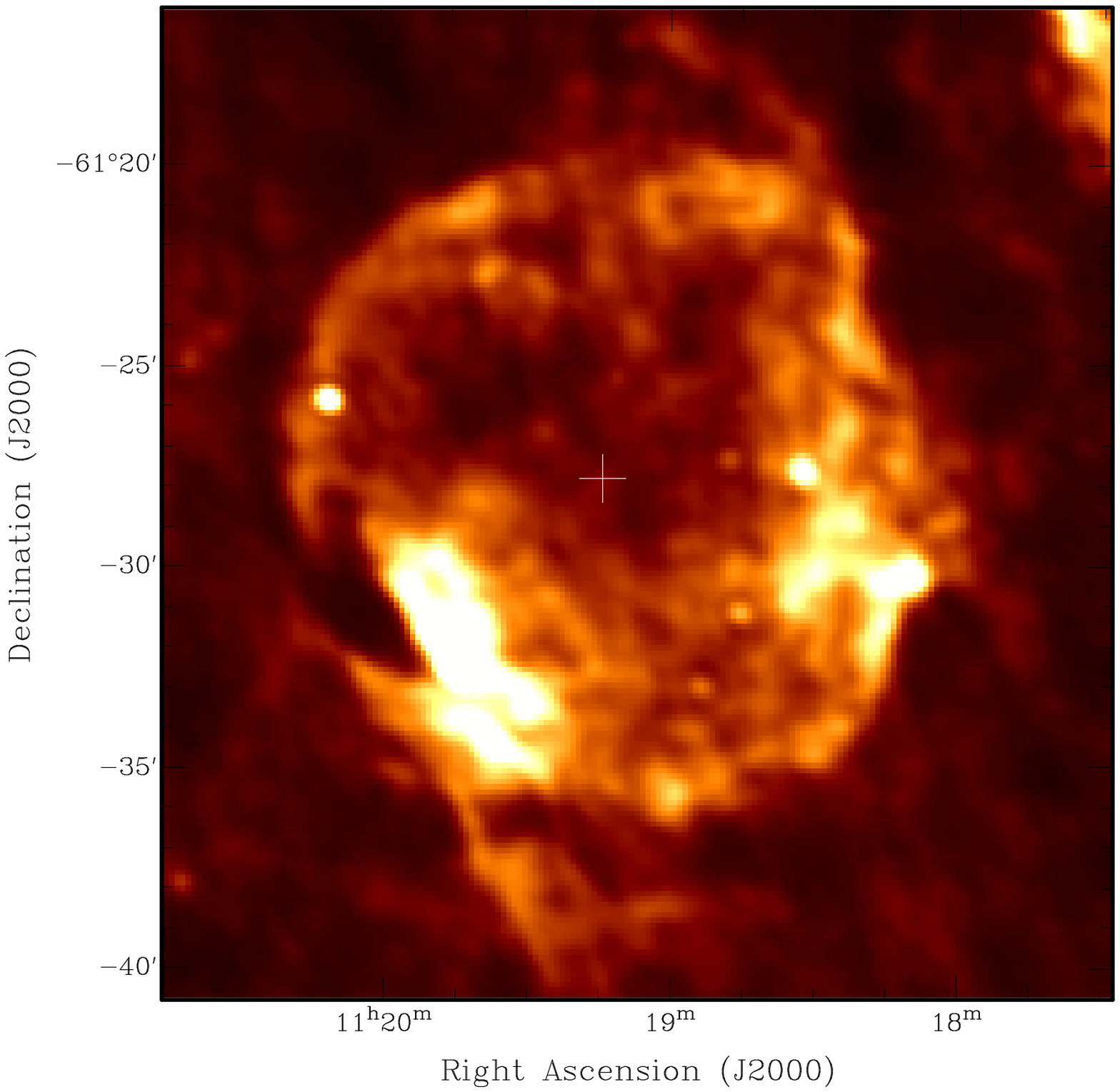}{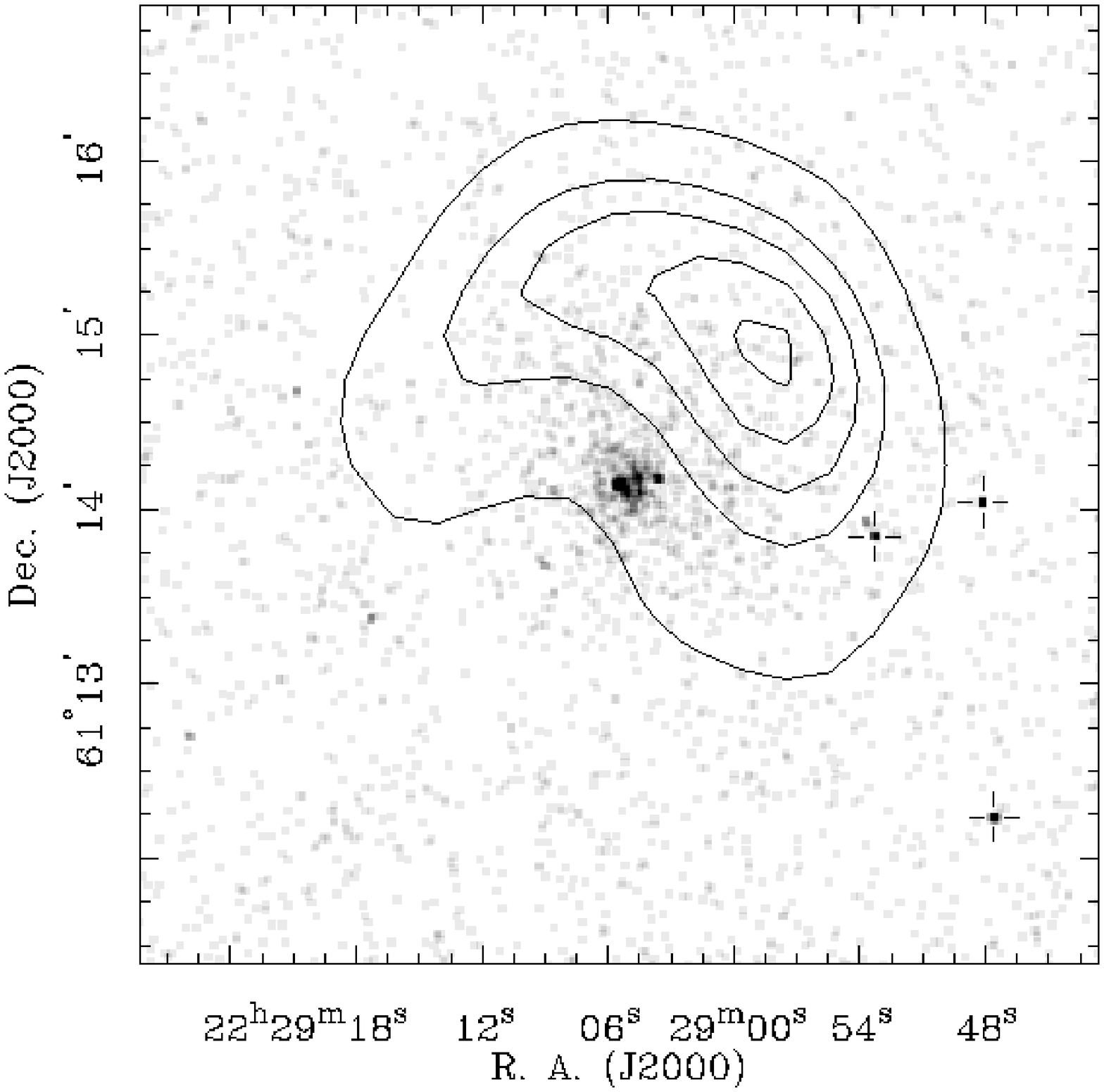}
\caption{(Left): PSR~J1119--6127 (cross; Camilo et al. 2000) centered
on the shell SNR~G292.2--0.5 (ATCA image; Crawford et al. 2001).
PSR~J1846--0258, with very similar spin parameters, was discovered
serendipitously in the composite SNR~Kes~75 by Gotthelf et al. (2000).
(Right):  VLA contours of 25\% polarized, flat spectrum source overlaid
on {\em Chandra\/} image of PWN~G106.6+2.9.  PSR~J2229+6114 is
coincident with the central point source (Halpern et al. 2001). }
\end{figure}

\vspace{-5mm}

\section{Deep Searches}
The certain existence of a PWN by definition points to the presence of
an energetic and reasonably young pulsar.  PWNe also have the virtue of
being compact (usually smaller than the primary beam size of a large
telescope at 1.4\,GHz, or 3--15\,arcmin).  Thus it is eminently
reasonable to search all PWNe {\em as deeply as possible}.  (Despite
arguments to the contrary there is no credible evidence suggesting that
young pulsars have higher radio luminosity than their older brethren
--- see Fig.~4.  If true, this would obviate the need to search deeper
than some threshold luminosity.)  Alas, not all that glitters is a
PWN.  (Some ``SNRs'' listed in the Green 2001 catalog are not even
clearly SNRs.)  However, with the {\em Chandra\/} X-ray Observatory and
its superb resolving power, one can for the first time often identify
unambiguously a PWN and its embedded pulsar even when pulsations are
not detected.  These are then targets worthy of the utmost efforts at
radio wavelengths in pursuit of pulsations, as I now demonstrate.

Energetic and young pulsars are the only substantial established class
of Galactic point $\gamma$-ray sources detected with EGRET on the {\em
CGRO\/} satellite.  Understanding the remaining unidentified EGRET
sources presents a considerable challenge owing to their large
positional uncertainties.  In an exhaustive multiwavelength study of
3EG~J2229+6122, Halpern et al. identified all but one of the X-ray
sources within its error box.  Its high ratio of X-ray-to-optical flux,
particularly when combined with a coincident polarized and
flat-spectrum radio source, strongly suggested a pulsar
interpretation.  Finally, a {\em Chandra\/} observation resolved a ring
of emission surrounding a point source (Fig.~1~right), much as for the
Vela PWN/pulsar, clearly indicating a PWN.  Pursuing this trail we
detected the pulsar in a 2\,hr observation using the 76\,m telescope at
Jodrell Bank, subsequent to which pulsations were detected in X-rays.
PSR~J2229+6114 has period $P = 51$\,ms, dispersion measure $\mbox{DM} =
205$\,cm$^{-3}$\,pc, characteristic age $\tau_c = P/2\dot P = 10$\,kyr,
and spin-down luminosity $\dot E = 4 \pi^2 I \dot P/P^3 = 2.2 \times
10^{37}$\,erg\,s$^{-1}$, where $I \equiv 10^{45}$\,g\,cm$^2$.  There is
no plausible alternative counterpart, and the pulsar is virtually
certainly the source of 3EG~J2229+6122's $\gamma$-rays (Halpern et al.
2001).  In all respects it appears to be similar to the Vela
pulsar/PWN, at $\sim 10$ times the distance.  It also has a low radio
luminosity: with a flux density at 1.4\,GHz of $S_{1400} = 0.25$\,mJy,
the luminosity is $L_{1400} \equiv S_{1400} d^2 \sim 2$\,mJy\,kpc$^2$.

\begin{figure}[h]
\plottwo{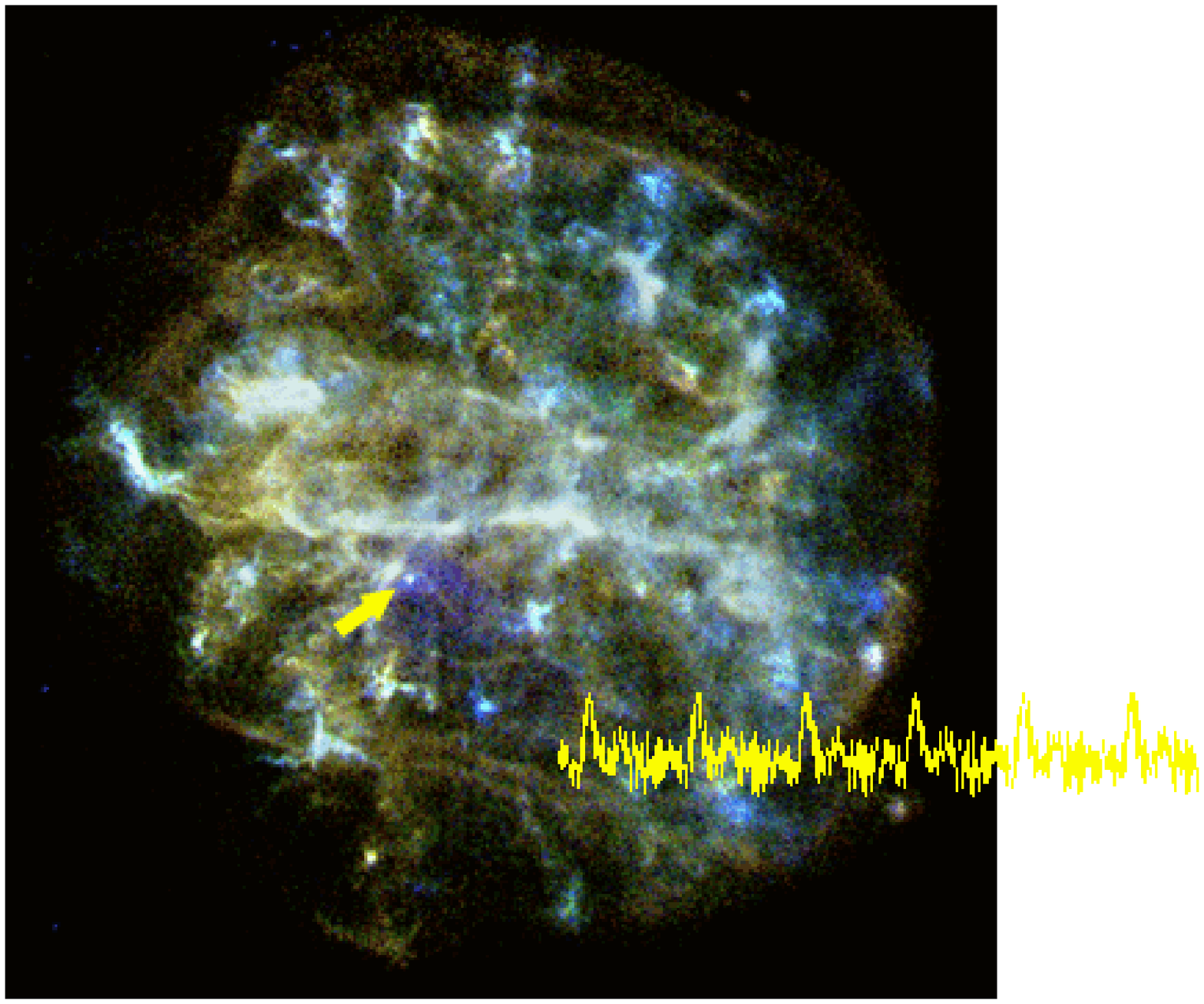}{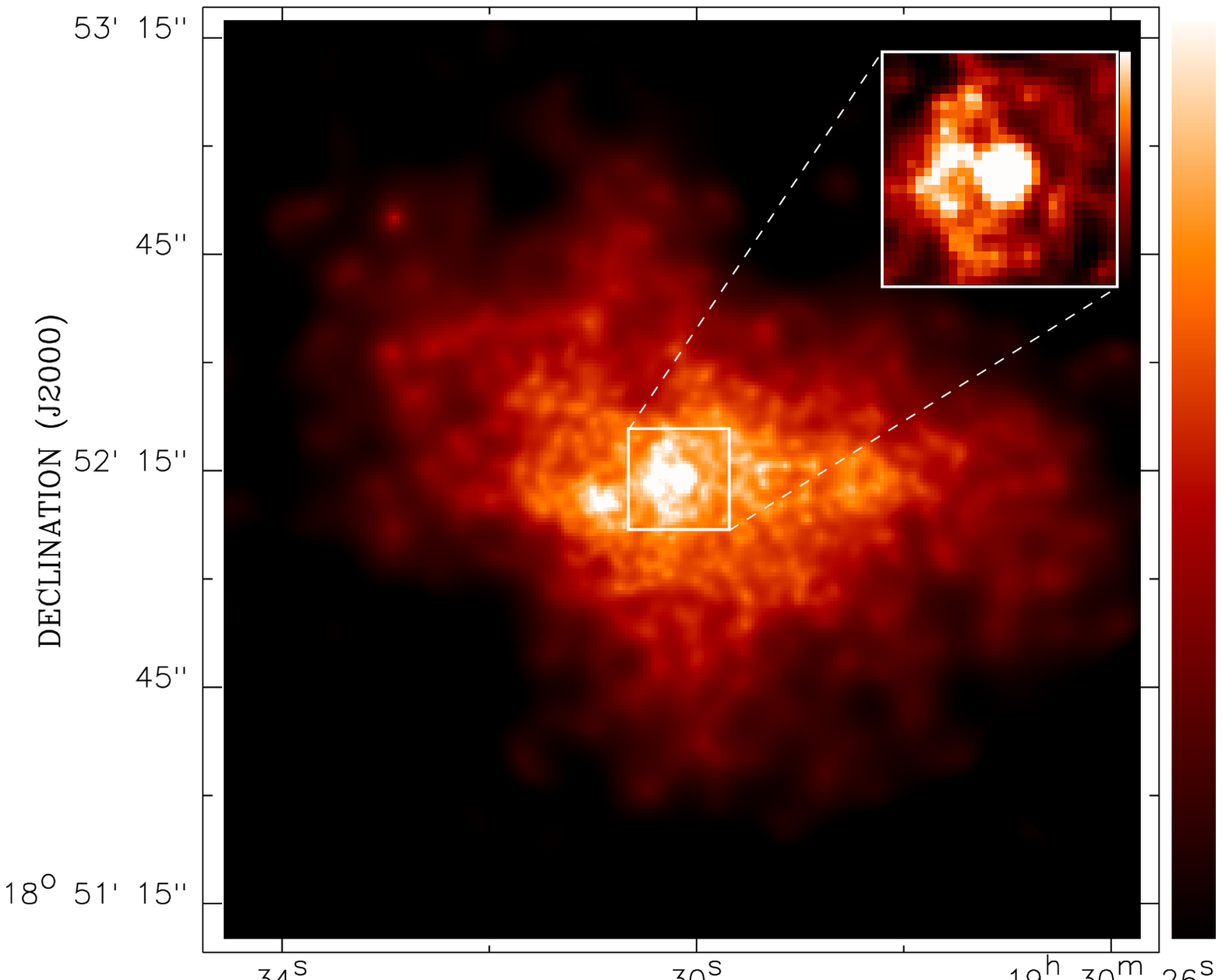}
\caption{(Left): {\em Chandra\/} image of SNR~G292.0+1.8 (Hughes et al.
2001) with position (arrow) and mean pulse profile of PSR~J1124--5916
(Camilo et al. 2002a) indicated.  (Right): {\em Chandra\/} image of
SNR~G54.1+0.3 (Lu et al. 2002).  PSR~J1930+1852 (Camilo et al. 2002c)
is coincident with the central point source seen in inset. }
\end{figure}

Figure~2 (left) shows a beautiful {\em Chandra\/} observation of the
oxygen-rich $\sim 1700$ year-old composite SNR~G292.0+1.8 (Hughes et
al. 2001).  Although not clear in this representation, the data clearly
resolve a $\sim 2\arcmin$ PWN within which is located a point source.
While this SNR had been searched previously (as had many of these
targets), the {\em Chandra\/} results encouraged new efforts.  In a
10\,hr integration at Parkes using the multibeam system, we detected
PSR~J1124--5916 (Camilo et al. 2002a), with $P = 135$\,ms, $\mbox{DM} =
330$\,cm$^{-3}$\,pc, $\tau_c = 2900$\,yr, and $\dot E = 1.2 \times
10^{37}$\,erg\,s$^{-1}$.  It is a very weak source, with $S_{1400} =
80\,\mu$Jy and $L_{1400} \sim 2$\,mJy\,kpc$^2$.  Figure~2 (right) once
again shows a spectacular {\em Chandra\/} image, of the ``Crab-like''
SNR~G54.1+0.3 (that is, a PWN, with no evidence for stellar ejecta or
thermal emission from interaction with the ambient ISM).  In a 3\,hr
observation at Arecibo using a frequency of 1.2\,GHz and bandwidth of
0.1\,GHz, we detected the pulsar coincident with the central point
source visible in the image.  PSR~J1930+1852 has spin parameters that
are virtually identical to those of PSR~J1124--5916, and is an even
fainter radio source (Camilo et al. 2002c).  X-ray pulsations were
detected in both of these objects subsequent to the radio discoveries.

3C58 is a Crab-like SNR thought to be the remnant of SN~1181.  After 20
years of searching at X-ray and radio wavelengths, PSR~J0205+6449 was
finally discovered with the {\em Chandra\/} and {\em RXTE\/} telescopes
(Murray et al. 2002).  Its spin parameters ($P = 65$\,ms, $\tau_c =
5400$\,yr, and $\dot E = 2.7 \times 10^{37}$\,erg\,s$^{-1}$) can be
reconciled with the 1181~AD explosion if the initial spin period was a
relatively large $\sim 60$\,ms.  Subsequent to this we detected the
pulsar with the Green Bank Telescope in individual 8\,hr integrations
at frequencies of 0.8 and 1.4\,GHz: it is the weakest of all (so far)
known young pulsars, with $S_{1400} = 50\,\mu$Jy and $L_{1400} \sim
0.5$\,mJy\,kpc$^2$, and has $\mbox{DM} = 141$\,cm$^{-3}$\,pc (Camilo et
al. 2002b).

\begin{figure}[h]
\plotone{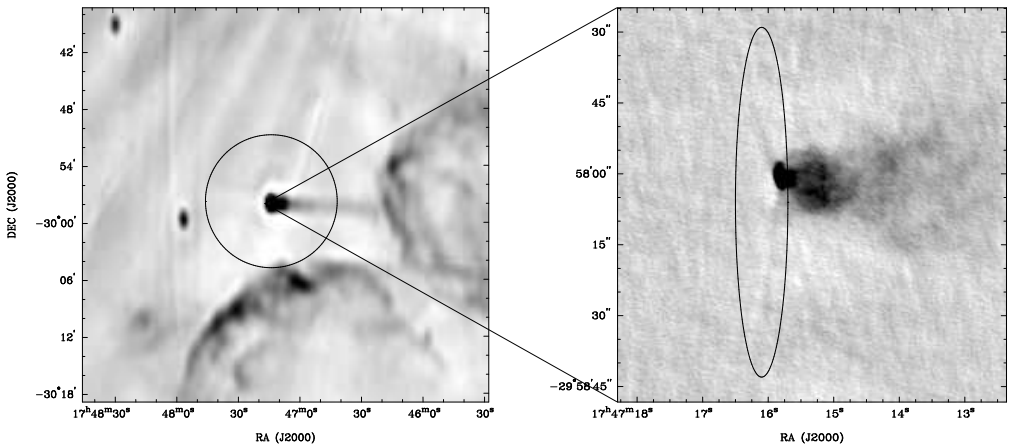}
\caption{(Left): Large scale view of G359.23--0.82 (``The Mouse''),
showing its bright head and long tail, and two unrelated SNRs (0.8\,GHz
MOST image).  (Right): Detailed view of the Mouse's head from VLA
observations at 8.4\,GHz.  The ellipse denotes the current positional
uncertainty of PSR~J1747--2958 (Camilo et al. 2002d).  }
\end{figure}

The fifth of our recent young pulsar discoveries is located near the
``Mouse'', an axisymmetric non-thermal nebula with a polarized tail,
long thought to be caused by a fast-moving pulsar (see Fig.~3).
PSR~J1747--2958, discovered in a 9\,hr observation at Parkes, has $P =
98$\,ms, $\mbox{DM} = 101$\,cm$^{-3}$\,pc, $\tau_c = 25$\,kyr, and
$\dot E = 2.5 \times 10^{36}$\,erg\,s$^{-1}$ (Camilo et al. 2002d).
Its distance, determined from the DM to be $\sim 2$\,kpc, implies a
luminosity $L_{1400} \sim 1$\,mJy\,kpc$^2$.  The probability of chance
coincidence between the Mouse's head and the pulsar, considering its
current positional uncertainty, is only $\sim 5 \times 10^{-5}$; the
distance constraints to both objects are consistent; and the ({\em
ROSAT\/}) X-ray energetics of the Mouse's head are compatible with
being powered by the pulsar.  We thus feel confident in associating
both objects: the Mouse is a PWN powered by PSR~J1747--2958, with the
morphology of the head expected to be shaped by ram-pressure balance
between the pulsar wind and the ISM.  Future measurements in this
system should be quite valuable to characterize the local ISM.

We have detected these five young pulsars associated with PWNe in the
course of searching a total of 20 PWNe.  We expect to search seven
additional PWNe shortly.  Also, we have made deep searches of five
other young neutron stars with no surrounding PWNe: the four well known
``radio-quiet neutron stars'' in SNRs G260.4--3.4, G266.2--1.2,
G296.5+10.0, and G332.4--0.4 (reaching luminosity limits $L_{1400} \sim
0.2$\,mJy\,kpc$^2$), and RX~J1836.2+5925 in 3EG~J1835+5918 (Halpern et
al. 2002).  We have no confirmed detections of pulsations from any of
these objects, although we have a very intriguing candidate from
G296.5+10.0 at the X-ray period observed by Pavlov et al. (2002).

\section{Discussion}

\vspace{-5mm}

\begin{figure}[h]
\plotone{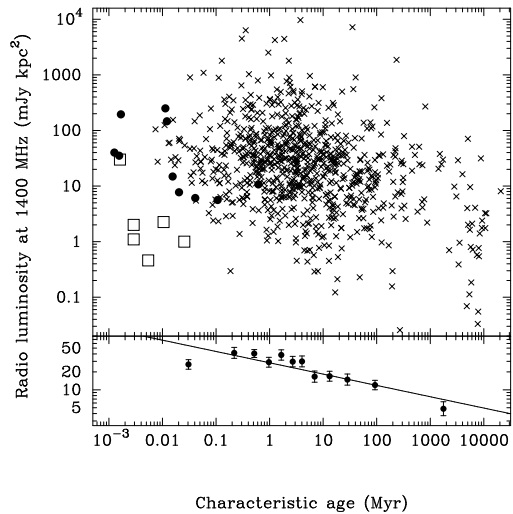}
\caption{$L_{1400}$ versus $\tau_c$ for 833 pulsars.  {\em Filled
circles\/} represent the nine Galactic radio pulsars known by 2000 to
be associated with SNRs, and PSR~B0540--69 in the LMC (at $\sim
200$\,mJy\,kpc$^2$ and 1600\,yr).  {\em Open squares\/} indicate the
recently discovered pulsars described here (PSR~J1119--6127 is at $\sim
30$\,mJy\,kpc$^2$). }
\end{figure}

\noindent The work summarized here indicates clearly that many young
pulsars beam towards the Earth, although what exact fraction is still
unclear.  What is abundantly clear is that many of them are very weak
radio sources (see Fig.~4).  When this project is completed, a careful
analysis of the sensitivity and selection effects relevant to the
search of the $\sim 25$ PWNe under study should yield useful
constraints on the combination of luminosity\footnote{It is important
to note that the radio luminosities discussed here are really
``pseudo-luminosities'': they assume that total luminosity is
proportional to the integrated flux density of the observed cut across
the radio beam.  A realistic discussion of actual luminosities depends
crucially on the generally unknown pulsar beam shape.} distribution and
beaming fraction of young pulsars.  In the meantime we may consider the
following: nearly half of all well-established PWNe already have
pulsations detected; the least luminous young radio pulsar known has
$L_{1400} \sim 0.5$\,mJy\,kpc$^2$; and few of the searches done to date
(including ones described here) reach this low level of luminosity.  I
am forced to conclude, optimistically, that from a purely observational
viewpoint nearly all remaining PWNe could contain a pulsar beamed at
us, and that several most certainly do.  The lesson is clear: one
should be discerning in choosing targets (often pointed to by {\em
Chandra\/} or {\em XMM\/}), and then (the wisdom of time-assignment
committees permitting) use {\em maximum effort in both time and
bandwidth\/} to pursue them.  Also, multi-path propagation cannot be
neglected for some distant objects: at $\sim 10$\,kpc along the
Galactic plane, some lines-of-sight have a predicted ``scattering
timescale'' of $\sim 20$\,ms at 1.4\,GHz.  In these cases, sensitive
$\sim 3$\,GHz systems with large bandwidth may be appropriate.  It is
hoped that several planned improvements at Parkes, Arecibo, and GBT
will be pushed to fruition soon, making possible even more sensitive
searches.

I conclude with one possibly important concern.  Nearly 200 of the 231
SNRs listed in the Green (2001) catalog are shells.  Could a
substantial number harbor pulsars?  I think yes: two of six associated
pulsars discovered in undirected searches are located in shell SNRs,
and have high $L_{1400} \sim 20$\,mJy\,kpc$^2$.  They also have spin
parameters rather different from those of most associated pulsars.
Thus we neglect these SNRs at our peril.  But how to go about searching
them?

\acknowledgements
It is a pleasure to thank all of my collaborators in this project, too
numerous to list here, and in particular Jules Halpern, Dunc Lorimer,
Bryan Gaensler, Dick Manchester, and Ingrid Stairs.  This research is
supported by NASA, NSF, and NRAO.


\vspace{-2mm}


\begin{references}
\reference Camilo, F. et al. 2000, \apj, 541, 367
\reference Camilo, F. et al. 2002a, \apj, 567, L71
\reference Camilo, F. et al. 2002b, \apj, 571, L41
\reference Camilo, F. et al. 2002c, \apj, 574, L71
\reference Camilo, F. et al. 2002d, \apj, 579, L25
\reference Crawford, F. et al. 2001, \apj, 554, 152
\reference Gorham, P.~W. et al. 1996, \apj, 458, 257
\reference Gotthelf, E.~V. et al. 2000, \apj, 542, L37
\reference Green, D.~A. 2001, http://www.mrao.cam.ac.uk/surveys/snrs/
\reference Halpern, J.~P. et al. 2001, \apj, 552, L125
\reference Halpern, J.~P. et al. 2002, \apj, 573, L41
\reference Hughes, J.~P. et al. 2001, \apj, 559, L153
\reference Kaspi, V.~M. et al. 1996, \aj, 111, 2028
\reference Kaspi, V.~M. \& Helfand, D.~J. 2002, in ASP Conf. Ser. 271,
 Neutron Stars in
 Supernova~Remnants,~ed.~P.~Slane~\&~B.~Gaensler~(San~Francisco:~ASP),~3
\reference Lorimer, D.~R., Lyne, A.~G., \& Camilo, F. 1998, \aap, 331, 1002
\reference Lu, F.~J. et al. 2002, \apj, 568, L49
\reference Manchester, R.~N. et al. 2001, \mnras, 328, 17
\reference Murray, S.~S. et al. 2002, \apj, 568, 226
\reference Pavlov, G.~G. et al. 2002, \apj, 569, L95
\end{references}
\end{document}